\documentclass[aps,prb,preprint,superscriptaddress]{revtex4}
\usepackage{graphicx}
\usepackage{bm}			
\begin{document}

\title{Charge-density-wave order takes over antiferromagnetism in Bi$_2$Sr$_{2-
x}$La$_x$CuO$_{6}$ superconductors}


\author{S. Kawasaki}
\thanks{These authors contributed equally.}
\affiliation{Department of Physics, Okayama University, Okayama 700-8530, Japan}

\author{Z. Li}
\thanks{These authors contributed equally.}
\affiliation{Institute of Physics, Chinese Academy of Sciences, and Beijing National Laboratory for Condensed Matter Physics,  Beijing 100190, China}

\author{M. Kitahashi}
\affiliation{Department of Physics, Okayama University, Okayama 700-8530, Japan}

\author{C. T. Lin}
\affiliation{Max-Planck-Institut fur Festkorperforschung, Heisenbergstrasse 1, D-70569 Stuttgart, Germany}

\author{P. L. Kuhns}
\affiliation{National High Magnetic Field Laboratory, Tallahassee, Florida 32310, USA}

\author{A. P. Reyes}
\affiliation{National High Magnetic Field Laboratory, Tallahassee, Florida 32310, USA}

\author{Guo-qing Zheng}
\thanks{To whom correspondence should be addressed; E-mail:  gqzheng123@gmail.com}
\affiliation{Department of Physics, Okayama University, Okayama 700-8530, Japan}
\affiliation{Institute of Physics, Chinese Academy of Sciences, and Beijing National Laboratory for Condensed Matter Physics,  Beijing 100190, China}


\begin{abstract}
Superconductivity appears in the cuprates when a spin order is destroyed, while the role of 
charge is less known. Recently, charge density wave (CDW) was found below the 
superconducting dome in YBa$_2$Cu$_3$O$_y$ when a high magnetic field is applied 
perpendicular to the CuO$_2$ plane, which was suggested to arise from incipient CDW in 
the vortex cores that becomes overlapped. Here, by $^{63}$Cu-nuclear magnetic 
resonance, we report the discovery of CDW induced by an in-plane field that does not 
create vortex cores in the plane, setting in above the dome in single-layered 
Bi$_2$Sr$_{2-x}$La$_x$CuO$_6$. 
 The onset temperature $T_{\rm CDW}$ takes over the antiferromagnetic order 
 temperature $T_{\rm N}$ beyond a critical doping level at which superconductivity starts 
 to emerge,  and  scales with  the psudogap temperature $T^{*}$.   These results provide 
 important insights into the relationship between spin order, CDW and the pseudogap, and 
 their connections to high-temperature superconductivity.
\end{abstract}

\pacs{}

\maketitle


\section{Introduction}

High transition-temperature ($T_{\rm c}$) superconductivity is obtained by doping carriers to destroy an antiferromagnetic (AF) spin ordered Mott insulating phase.
Although it is generally believed that the interaction responsible for the spin order is 
important for the superconductivity \cite{LeeNagaosaWen}, 
the nature of the  normal state  is still unclear \cite{Uchidareview,Fradkin}. For example, in 
the region with low carrier concentration $p$ (0 $< p <$ 0.2), a pseudogap state emerges 
where   partial  density of states (DOS) is lost below a characteristic temperature $T^*$   
well above  $T_{\rm c}$  \cite{Timusk} or even $T_{\rm N}$ \cite{Kawasaki}. Although 
the nature of the strange metallic state is still under debate, it is likely connected to both 
spin and change fluctuations or even orders. 
 In fact, experimental progress suggests that the spin and charge degrees of freedom are highly entangled. 

For example,  a  striped spin/charge  order was found around $x$$\sim$1/8 in 
La$_{1.6-x}$Nd$_{0.4}$Sr$_x$CuO$_4$ (LSCO) two decades ago \cite{Tranquada}.  
More recently, various forms of  charge order were reported in many other systems. 
Scanning tunneling microscopy (STM) in Bi$_2$Sr$_2$CaCu$_2$O$_{8+\delta}$ found a modulation of local DOS in the vortex cores where superconductivity is destroyed \cite{Hoffman}, which was interpreted as due to  halos of incipient CDW localized within the cores \cite{Sachdev,KivelsonLee}.
Resonant elastic and inelastic x-ray spectroscopy (RXS) measurements found a short-range CDW with ordering vectors along the in-plane Cu-O bond directions, $\bf q$ =($\sim$0.3, 0) and (0, $\sim$0.3). The correlation length is $\xi_{a,b} \sim$ 50 \AA \  for YBa$_2$Cu$_3$O$_{7-y}$ (YBCO)  \cite{Ghiringhelli,ChangNatPhys} and $\xi_{a,b} \sim$ 20 \AA \ for the other systems 
\cite{CominBi2201,SilvaNetoBi2212,PengYY,HashimotoBi2212,Tabis}. Quite recently, it was suggested by $^{17}$O nuclear magnetic resonance (NMR) in YBCO that such CDW is of static origin \cite{WuONMR}.
In Bi$_2$Sr$_{2-x}$La$_x$CuO$_{6+\delta}$ (Bi2201), 
most intriguingly, the onset temperature  of the short-range CDW was found 
 to coincide \cite{CominBi2201,PengYY} with  $T^*$  that is far above $T_{\rm c}$ or $T_{\rm N}$ \cite{Zheng,Kawasaki}.

Application of a high magnetic field is useful to diagnose the interplay between various orders in the cuprates. When a high magnetic field is applied   perpendicular to the CuO$_2$ plane,  superconductivity  can be  suppressed substantially.   In YBCO, 
  $^{63}$Cu  NMR at $H$ = 28.5 T revealed  a long-range charge density modulation   
  perpendicular to the CuO-chain  in the sample with $p$ = 0.108 \cite{WuNature}.
  RXS also indicated  that a high field induces a correlation along the CuO-chain  direction and modifies the coupling between CuO$_2$ bilayers, thus causes a three-dimensional CDW \cite{Nojiri,Changhighfield}. These observations  are consistent with early discovery of a Fermi-surface reconstruction   by  quantum oscillations \cite{Leyraud1} and  a  recent report of a thermodynamic phase transition \cite{LeBoeuf}.  

These findings have arisen much interests, but the origin of the CDW and its connection to 
superconductivity is yet unknown. As  the  long-range CDW onsets below  $T_{\rm 
c}(H=0)$ and only emerges when the field is applied perpendicular to the CuO$_2$ plane, 
 a wide-spread speculation  is that it is due to incipient CDW in the vortex cores 
 \cite{Hoffman} that becomes overlapped  as the field gets stronger 
 \cite{ChangNatPhys,SilvaNetoBi2212,WuVortex}. 
In fact, a  field as large as  28.5 T applied in the CuO$_2$ plane of YBCO did not bring 
about any long-range CDW \cite{WuNature}.
 Also, the role of the CuO chain is unclear; in Bi$_2$Sr$_2$CaCu$_2$O$_{8+\delta}$ 
 without a CuO chain, no long-range CDW was found \cite{CurroBi2212}.

 In order to clarify the relationship between the intertwined   AF spin order, CDW, pseudogap and superconductivity, we apply high magnetic fields up to 42.5 T parallel to the Cu-O bond direction ($H \parallel $ $a$ or $b$ axis) in Bi$_2$Sr$_{2-x}$La$_x$CuO$_{6+\delta}$ where the pseudogap spans from the parent AF insulating phase to the overdoped superconducting regime \cite{Kawasaki,Zheng}. This material has no CuO chain and the application of an in-plane field does not create vortex cores in the CuO$_2$  plane.
Surprisingly,  
we discover a long-range CDW that emerges far above the superconducting dome for 
$H_{\parallel}$ $>$ 10 T. We find that such CDW order becomes the successor of the AF 
order beyond $p$ = 0.107 at which superconductivity starts to emerge. The $T_{\rm 
CDW}$  takes over  $T_{\rm N}$, but disappears well before the pseudogap closes.
Our results indicate that  CDW can be well disentangled from other orders.

\section{Results}

 \noindent     
\textbf{Evidence for a field-induced CDW in underdoped Bi2201.}
 Figures 1a-1d shows the $^{63}$Cu-NMR satellite  (3/2$
 \leftrightarrow$1/2 transition) lines for four compounds of Bi$_2$Sr$_{2-x}$La$_x$CuO$_{6+\delta}$ at two representative temperatures  at $H_{||}$ = 14.5 or 20.1 T.  
As seen in Fig. 1a, no change between $T$ = 100 K and 4.2 K is observed for the optimally-doped compound ($p$ = 0.162). However, the spectrum is broadened at $T$ = 4.2 K for $p$ = 0.135 (Fig. 1b), and  a splitting $\pm\delta{f}$ of the spectrum is observed at $T$ = 4.2 K  for $p$ = 0.128 and 0.114,  as seen in Figs. 1c and 1d. The  spectra at $T$ = 4.2 K for $p$ = 0.128 and 0.114 can be  reproduced by a sum of two Gaussian functions. It is noted that at low fields  below 10 T, the spectrum shows no appreciable temperature dependence in the whole temperature range.
The NMR line splitting indicates a long-range order, as it measures an assemble of nuclear 
spins over the sample.

Figure 1e shows the field evolution of $\delta f$  for  $p$ = 0.114. The $\delta f$ grows 
steeply at $H_{||}$ = 10.4 T and saturates above $H_{||}$ $\sim$ 14.5 T. Figure 1f 
shows the temperature dependence of $\delta f$ for $p$ = 0.114 under various fields. The 
$\delta{f}$ grows rapidly below $T$ $\sim$ 30,  55, and 60 K at $H_{||}$ = 11, 13, and 
above  14.5 T, respectively. These results indicate that a field-induced phase transition 
occurs in the underdoped Bi2201. The results are qualitatively similar to the results found in 
YBCO where the same transition line splits into two peaks due to the spatial modulation of 
the NQR frequency $\nu_{\rm Q}$ \cite{WuNature}.

Next we show that  the field-induced phase transition is due to a charge order, but not spin 
order.  Figures 2a, b, respectively, shows the temperature dependence of the satellite 
and the center lines  for the sample with the lowest doping  $p$ = 0.114. A spectrum 
broadening is also found in the center line (1/2$ \leftrightarrow$-1/2 transition) at $T$ = 
4.2 K,  but it is much smaller than the satellite line. Figure 2c shows the temperature 
dependence of the NMR intensity obtained by integrating  the spectrum at each 
temperature. The intensity has no anomaly above $T_{\rm c}$, indicating that there is no 
spin order. For antiferromagnetic insulator $p$ = 0.107 ($x$ = 0.08), the intensity 
decreases below $T_{\rm N}$ = 66 K because an internal field $H_{\rm int}$ = 7 T shifts 
the peak frequency far away \cite{Kawasaki}. Furthermore, a possibility of  striped phase 
formation leading to a wipe out effect found in LSCO \cite{Hunt} can also be ruled out. It is 
noted that the possibility of a field-induced spin-density-wave order has already been ruled 
out previously for $p$ = 0.162 \cite{Mei}.

Therefore, the splitting of the satellite line (Fig. 2a) and the broadening of the center line (Fig. 2b) is due to a distribution of the Knight shift $K_{\parallel}\pm\delta K_{\parallel}$ and the NQR frequency as $\nu_{\rm Q}\pm\delta\nu$ as observed in YBCO. Furthermore, the splitting $\delta f_{\rm satellite}$ = 1.22 MHz is much larger than $\delta f_{\rm center}$ = 0.271 MHz, indicating that the $\nu_{\rm Q}$ change is the main contributor to the observed line splitting.  By a simple calculation (see Supplementary Note 1), we find that $\delta$$K_{\parallel}\sim0.05\pm0.01\%$ and $\delta\nu$ =  2.5 $\pm$ 0.2 MHz can reproduce both the satellite and the center lines at the same time (shaded areas in Figs. 2a and 2b).
The relation $\nu_{\rm Q}$ = 22.0 + 39.6 $p$ (see Supplementary Fig. 1) then yields a   hole-concentration distribution $\delta p \sim 0.06 \pm 0.01$ at the Cu site.
 Since there is no spin order here (Fig. 2c) as mentioned already, the splitting of the satellite line $\delta f$ ($\propto\delta p$) indicates a field-induced long-range charge distribution, {\it i.e.}, a formation of CDW at low temperature in underdoped Bi2201.   

Figure 2d, e shows the temperature dependence of the nuclear spin-lattice relaxation 
rate divided by $T$,  $1/T_1T$  and spin-spin relaxation rate $1/T_2$ for $p$ = 0.114 
obtained at two different fields. At $H$ = 9 T,  both  quantities decrease monotonically  
below $T^*$ = 230 K.   At  $H$ = 13 T, however, a pronounced peak was found in 
$1/T_1T$ at $T_{\rm CDW}$ =55 K. Such a peak in $1/T_1T$ is a characteristic of a CDW 
order \cite{SrPt2As2}. The $1/T_2$ also shows a sharp decrease at $T_{\rm CDW}$ =55 
K.  These results provide further evidence for a  field-induced CDW phase transition.

\vspace{0.5cm}
\noindent   
\textbf{$H$-$T$ phase diagram for underdoped Bi2201.} To obtain the CDW onset  
temperature ($T_{\rm CDW}$)  and the threshold field ($H_{\rm CDW}$) for $p$ = 
0.114, we study the temperature dependence of the NMR spectra at various magnetic fields 
(Figs. 1e, f).

Figure 3 shows the $H$-$T$ phase diagram for $p$ = 0.114. 
Remarkably, the long-range  CDW state in Bi2201 emerges at a temperature far above $T_{\rm c}$, in contrast to that in YBCO where  CDW appears below $T_{\rm c}(H=0)$ \cite{WuNature,WuVortex}.

\vspace{0.5cm}

 \noindent   
  \textbf{Relationship between CDW and superconductivity.}  Figure 4a, b shows the 
  $H$-dependence and $T$-dependence of the satellite splitting $\delta{f}$ which allow us 
  to obtain 
  $H_{\rm CDW}$ and $T_{\rm CDW}$ for various doping levels. Figure 5 shows the hole 
  concentration dependence of $H_{\rm c2}$, $H_{\rm CDW}$, and $T_{\rm CDW}$. 
  The $H_{\rm c2} \sim$ 60 T for $p$ = 0.162 (see Supplementary Fig. 5) decreases with 
  decreasing doping level but increasing again at $p$ = 0.114. Although the previous 
  Nernst effect study on three Bi2201 samples  ($p$ = 0.12, 0.16, and  0.19) did not take a 
  closer look into the doping range as we did here \cite{ChangHc2}, 
our result is consistent with the results of  YBCO \cite{WuVortex} and La$_{1.8-x}$Eu$_{0.2}$Sr$_x$CuO$_4$ \cite{ChangHc2}.

 The  $H_{\rm CDW}$ is slightly lower than that in YBCO, suggesting that  CDW has a similar energy scale across different class of cuprates. 
 However, the relationship between $H_{\rm CDW}$ and  $H_{\rm c2}$ is completely 
 different from that seen in YBCO where $H_{\rm CDW}$  scaled with $H_{\rm c2}$. 
 Namely, $H_{\rm CDW}$ was the lowest at the doping concentration where  $H_{\rm 
 c2}$ was the smallest there \cite{WuVortex}, which  led to the suggestion that  CDW  can 
 only be seen when the superconducting state is suppressed as the vortex cores become 
 overlapped. In the present case, however, no vortex cores are created in the CuO$_2$ 
 plane. In fact, $H_{\rm CDW}$ and $T_{\rm CDW}$ are 
 more related with doping concentration itself as can be seen in Fig. 5, rather than with 
 $H_{\rm c2}$. Namely, the long-range CDW order is induced more easily closer to the AF 
 phase boundary.

\section{Discussion}

  In this section, we  discuss possible CDW form and the implication of the phase diagram 
  we found. First, the result can be understood by an incommensurate 1-dimensional (1D) 
  long-range 
CDW as follows, as the situation is similar to that observed at the in-plane Cu$^{\rm 
2F}$-site located below the oxygen-filled CuO chain in YBCO \cite{WuNature}. For an 
unidirectional CDW state, the wave modulation causes a spatial distribution in the electric 
field gradient (EFG) and thus the NQR frequency so that $\nu$ = $\nu_{\rm 
Q}$+$\delta\nu$$\cos$$(\phi(X))$  \cite{Blinc,SrPt2As2}, where $X$ (= $a$ or $b$ 
axis) is the modulation direction.   The NMR  spectral intensity ($I(\nu)$)  depends on the 
spatial variation of $\phi(X)$ as  $I(\nu)=1/(2\pi d\nu/d\phi)$.  For an  incommensurate 
order, $\phi(X)$ is proportional to $X$, so that the NMR  spectrum shows an edge 
singularity at $\nu$ = $\nu_{\rm Q}$ $\pm$ $\delta\nu$, as $I(\nu)=1/(2\pi \delta\nu 
\sqrt{1-((\nu-\nu_Q)/\delta\nu)^2})$ \cite{Blinc,SrPt2As2}. By convoluting a 
broadening function, a two-peak structure can be reproduced. In such case, the quantity 
2$\delta\nu$ corresponds to the CDW order parameter \cite{SrPt2As2}.
 We emphasize that  the value of $\delta p$  is twice larger than that observed in YBCO \cite{WuNature}, indicating that a larger CDW amplitude is realized in Bi2201. This difference may arise from the different crystal structure between the two systems. YBCO is a bi-layer system while Bi2201 is  single-layered. When CDW has a different phase between two CuO$_2$ planes, the ordering effect would be weakened or even canceled out.  

It has been known that magnetic field works to induce a correlation along the CuO-chain direction and modifies the coupling between CuO$_2$ bilayers in YBCO \cite{Nojiri,Changhighfield}. In the present case, the short-range CDW at zero magnetic field \cite{CominBi2201,PengYY} could also be modified by $H_{\parallel} > $ 10 T to be a long ranged 1D CDW along the Cu-O direction. 

Second, how about a 2D-CDW case?  Recent resonant X-ray scattering measurement on 
Bi2201 $p$ $\sim$ 0.11 found a perfect 2D but local ($\xi \sim$ 20 $\AA$) CDW 
formation along the Cu-O bond direction with the wave vector ($Q^*$, 0) and (0, $Q^*$) 
($Q^*$ $\sim$ 0.26) $\cite{CominNatMat,PengYY}$. On the other hand, STM 
measurement suggested a commensurate density-wave with the ordering vectors 
$\bf{q}_{\rm DW}$ = (0.25, 0) and (0, 0.25) $\cite{DavisPNAS}$. In either case, if  
such a local CDW becomes long ranged with the same ordering vectors, 
a spatial distribution of the NQR frequency can be written as $\nu$ = $\nu_{\rm 
Q}$+$\delta\nu_X$$\cos$$(\phi(X))$+$\delta\nu_Y$$\cos$$(\phi(Y))$, where $X$ = 
$a$-axis and $Y$ = $b$-axis are the modulation directions $\cite{Blinc}$. As suggested by 
RXS and STM measurements, when the CDW amplitudes are equivalent $\delta\nu_X$ = 
$\delta\nu_Y$, such CDWs yield $I(\nu) \sim -ln[(\nu-\nu_{\rm 
Q})/\delta\nu]/\delta\nu$ $\cite{Blinc}$. In this case,  the logarithmic singularity 
appears at   $\delta\nu$ = 0 $\cite{Blinc}$, which is different from the 1D case. It is 
obvious that we can not explain our experimental results with such 2D CDW. However,  if 
the amplitude for each directions is different, 
 $\delta\nu_{X(Y)} \gg \delta\nu_{Y(X)}$, two edge-singularities will appear $\cite{Blinc}$. It is also interesting to note that if there are  CDW  domains with modulations  $\nu_X$ = $\nu_{\rm Q}$+$\delta\nu_X$$\cos$$(\phi(X))$ and $\nu_Y$ = $\nu_{\rm Q}$+$\delta\nu_Y$$\cos$$(\phi(Y))$  in each domain, the NMR lineshape will be the same as in the 1D case.

Third, a 3D CDW will not show two edge singularities in $\nu_{\rm Q}$ distribution in any 
case $\cite{Blinc}$. Therefore, we may exclude the possibility of a 3D CDW because, being 
different from the bilayer YBCO, the long distance between CuO$_2$ planes produces no 
CDW correlation along the $c$-axis in Bi2201 \cite{PengYY} and in our case the magnetic 
field is applied parallel to the CuO$_2$ plane.

 We now show  in Fig. 6
 how the long-range CDW emerges as the  magnetic field is increased. 
  As seen in the $H$-$p$ plane at $T$ = 4.2 K, the field-induced CDW emerges for $H_{\parallel}$ $>$ 10 T in the underdoped regime. At such high fields, 
  upon increasing doping, the AF  state with $T_{\rm N}$ = 66 K  at $p$ = 0.107 changes  to a CDW ordered state with $T_{\rm CDW}\sim$ 60 K at $p$ = 0.114. 
Upon further  doping  to $p$ = 0.162 where the pseudogap persists; however, the CDW 
order disappears. Although a detailed analysis  is difficult, a similar field-induced CDW is 
also  found when the magnetic field is applied perpendicular to the CuO$_2$ plane (see 
Supplementary Fig. 6).  
Figure 7 compares the phase diagram with that for YBCO.
In YBCO,  the short-range CDW sets in far below $T^*$ and the field-induced CDW 
(FICDW) occurs inside the superconducting dome forming a dome-like shape 
\cite{WuNature}, while in Bi2201  the short-range CDW sets in right at $T^*$ and the 
FICDW emerges  far above the superconducting dome and coexists with superconductivity.

Figure 6 reveals several important things. First and most remarkably, down-shifting the $T^*$ curve in temperature coincides with the $T_{\rm CDW}$  curve. As can be seen more directly in Fig. 8, $T_{\rm CDW}$ scales with  $T^*$.
 This may suggest that the psudogap is a fluctuating form of the long-range order found in this work, but more work is needed.  
Very recently, a polar Kerr effect \cite{ZXShen}
and an optical rotational anisotropy measurements \cite{Hsieh} suggested that a possible 
phase transition takes place at  $T^*$. However, we note that the probes used are ultrafast 
in time scale. In NMR measurements, the time scale is in the 10$^{-8}$ s range which is 
much slower than the optical measurements \cite{ZXShen,Hsieh}, and it is reasonable that 
$T^*$ is seen as a fluctuating crossover temperature. 
Second, $T_{\rm N}$  is succeeded  by $T_{\rm CDW}$ beyond a critical doping level at which superconductivity emerges, pointing to the important role of charge degree of freedom in high-temperature superconductivity.
However, the detailed evolution from AF to CDW order under high magnetic fields is unclear at the moment. It is a future task to clarify whether the evolution is  a  first-order phase transition or not.  The result also calls for further  scrutinies of the AF insulating phase. In fact,  the entanglement of the spin and charge freedoms \cite{Fradkin,TKLee} was recently found to occur  already in the  insulating phase itself \cite{WangYY}. In any case,
our results show that CDW order is another outstanding quantum phenomenon that needs to be addressed on the same footing as the AF spin order. Finally, the first demonstration of the ability of using an in-plane field to tune the electronic state should stimulate more works that will eventually help to solve the problem of high-$T_{\rm c}$ superconductivity.

 \section*{Methods}
\noindent
 \textbf{Samples.} The single crystals of Bi$_2$Sr$_{2-x}$La$_x$CuO$_{6+\delta}$ 
 (Bi2201, $p$ = 0.114 ($x$ = 0.75), 0.128 (0.65), 0.135 (0.60), and 0.162 (0.40)) were 
 grown by the traveling solvent floating zone method \cite{Liang,Peng}. The hole 
 concentration ($p$) were estimated previously \cite{Ono}. Small and thin single-crystal 
 platelets, typically sized up to 2 mm-2 mm-0.1 mm, cleaved from an as-grown ingot, were 
 used. The in-plane Cu-O bond direction ($a$ or $b$ axis) was determined by Laue 
 reflection. $T_{\rm c}(H)$ is defined at the onset temperature of diamagnetism observed 
 by ac-susceptibility measurement using NMR coil. $H_{\rm c2}$ is determined by fitting 
 $T_{\rm c}(H)$ to the WHH formula \cite{WHH}.
  
 \noindent 
 \textbf{NMR.} The $^{63}$Cu-NMR spectra were taken by sweeping the rf frequency at a 
 fixed field below $H$ = 15 T but they were taken by sweeping the field at a fixed frequency 
 above $H$ = 15 T. Measurements at $H$ = 14.5 T were conducted at Institute of Physics, 
 CAS, Beijing, and those below $H$ = 14.5 T were conducted at Okayama University. High 
 magnetic fields above $H$ = 15 T are generated by the Hybrid magnet in the National High 
 Magnetic Field Laboratory, Tallahassee, Florida. For $^{63,65}$Cu, the nuclear spin 
 Hamiltonian is expressed as the sum of the Zeeman and nuclear quadrupole interaction 
 terms, $\mathcal{H}$ = $\mathcal{H}_{\rm z} + \mathcal{H}_{\rm Q}$ = 
 $-^{63,65}\gamma\hbar{\bf I}\cdot{\bf{H}_{0}} (1+K) + (h \nu_{\rm 
 Q}/6)[3{I_z}^2-I(I+1)+\eta({I_x}^2-{I_y}^2)]$, where $^{63}\gamma$ = 11.285 
 MHz/T and $^{65}\gamma$ = 12.089 MHz/T, $K$ is the Knight shift, and $I$ = 3/2 is 
 the $^{63,65}$Cu nuclear spin. The NQR frequency $\nu_{\rm Q}$ and the asymmetry 
 parameter $\eta$ are defined as $\nu_{\rm Q}$  = $\frac{3eQV_{zz}}{2I(2I-1)h}$, 
 $\eta$ $=$ $\frac{V_{xx} - V_{yy}}{V_{zz}}$, with $Q$ and $V_{\alpha \beta}$ 
 being the nuclear quadrupole moment and the electric field gradient (EFG) tensor 
 \cite{abragam}. The principal axis $z$ of the EFG is along the $c$ axis and $\eta$ =  0 
 \cite{ZhengNuQ}. Due to $\mathcal{H}_{\rm Q}$, one obtains the NMR center line and 
 the two satellite transition lines between $|m\rangle$ and $|m-1\rangle$, ($m$ =3/2, 
 1/2, -1/2), at $\nu_{m\leftrightarrow m-1} = ^{63,65}\gamma H_0(1+K) + 
 (\nu_{\rm Q}/2)(3\cos^2\theta -1)(m-1/2)$ + second-order correction.  The second 
 term of the right side is the first order term in the presence of quadrupole interaction.  
 Here, $\theta$ is the angle between $\bf{H}$ and EFG.  
 The  $T_1$ and $T_2$ were measured at the frequencies in the center peak ($m = 1/2 
 \leftrightarrow$ - 1/2 transition). The $T_1$ values were measured by using a single 
 saturating pulse and were determined by standard fits to the recovery curve of the nuclear 
 magnetization to the theoretical function for the nuclear spin $I$ = 3/2 \cite{Zheng}. The 
 $T_2$ values 
 were obtained by fits to the spin-echo decay curve of the nuclear magnetization $I(t)$ to 
 $I(t)=I(0)\exp{(-2t/T_2)}$ \cite{Kawasaki}.

\noindent
\section*{Data availability}

The data that support the findings of this study are available on reasonable request. Correspondence and requests for materials should be addressed to G.-q.Z.

\section*{Acknowledgments}
We thank D.-H. Lee, S. Uchida, L. Taillefer, T. K. Lee, M.-H. Julien and  S. Onari for useful discussion, and S. Maeda and D. Kamijima for experimental assistance. A portion of this work was performed at National High Magnetic Field Laboratory, which is supported by NSF 
Cooperative Agreement No. DMR-1157490 and the State of Florida. Support  by research 
grants from Japan MEXT (No. 25400374 and 16H04016), China NSF (No. 11634015), and 
MOST of China (No. 2016YFA0300502 and No. 2015CB921304) is acknowledged.

 \section*{Authors contributions}
  G.-q.Z planned the project. C.T.L synthesized  the  single crystals. 
 S.K,  Z.L, M.K, P.L.K,  A.P.R and  G.-q.Z performed  NMR measurements. G.-q.Z wrote the manuscript with inputs from  S.K.  All authors discussed the results and interpretation.

\clearpage

\begin{figure}
      \begin{center}
      \includegraphics[width=15cm]{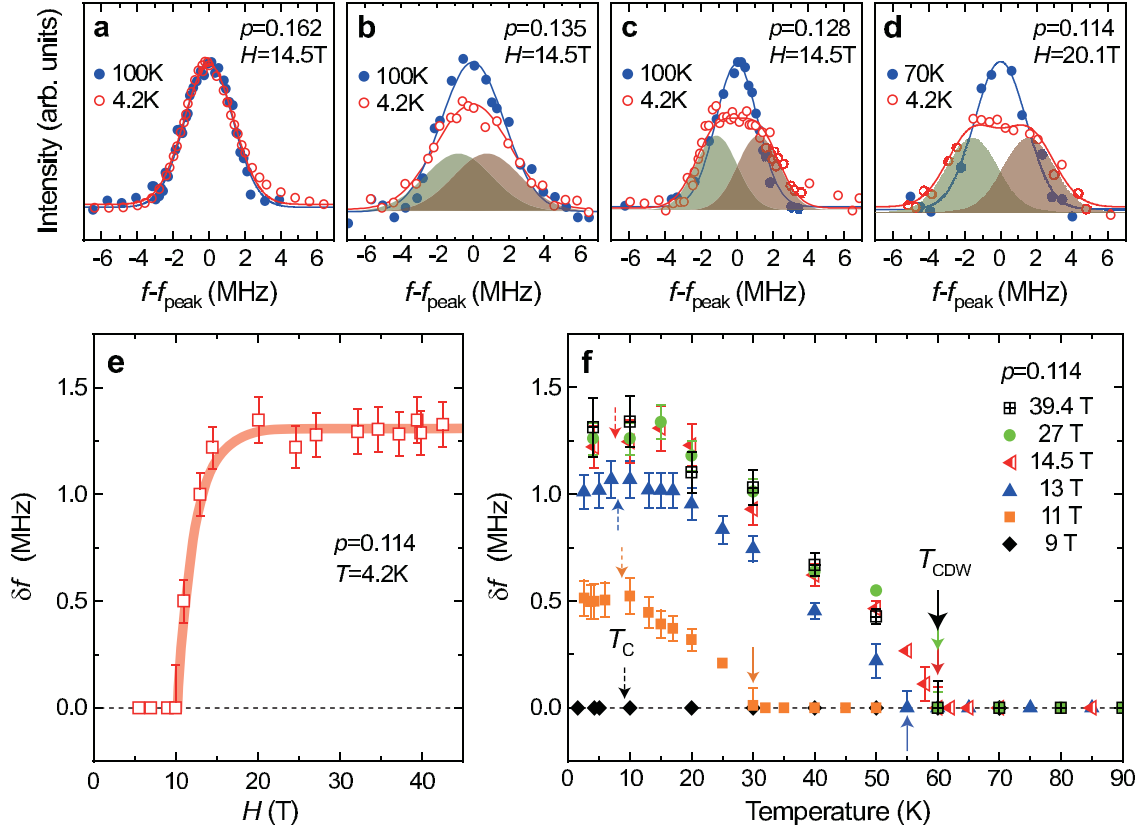}
      \end{center}
     
      \caption{Field and temperature evolution of NMR satellite line for underdoped 
      Bi2201. (a-d) The NMR satellite  (3/2$ \leftrightarrow$1/2 transition) lines for  
      Bi$_2$Sr$_{2-x}$La$_x$CuO$_{6+\delta}$ with  $x$ = 0.4 ($p$ = 0.162), $x$ = 
      0.6 ($p$ = 0.135), $x$ = 0.65 ($p$ = 0.128), and  $x$ = 0.75 ($p$ = 0.114). The 
      curves  for $p =$ 0.135, 0.128 and 0.114 at $T$ = 4.2 K are the sum of two Gaussian 
      functions (shaded area).  $f_{\rm peak}$ is the peak frequency. The intensity is 
      normalized by the area of the spectrum. (e) Field evolution of the line splitting $\delta 
      f$ for $p$ = 0.114 at $T$ = 4.2 K. (f) Temperature evolution of $\delta f$ for $p$ = 
      0.114 under various  fields.  The solid arrows indicate $T_{\rm CDW}$. 
      The dotted arrows indicate $T_{\rm c}(H)$. Error bars represent the uncertainty in 
      estimating $\delta{f}$. }   
        
      \label{f1}
      \end{figure}

       \begin{figure}
       \begin{center}
        \includegraphics[width=15cm]{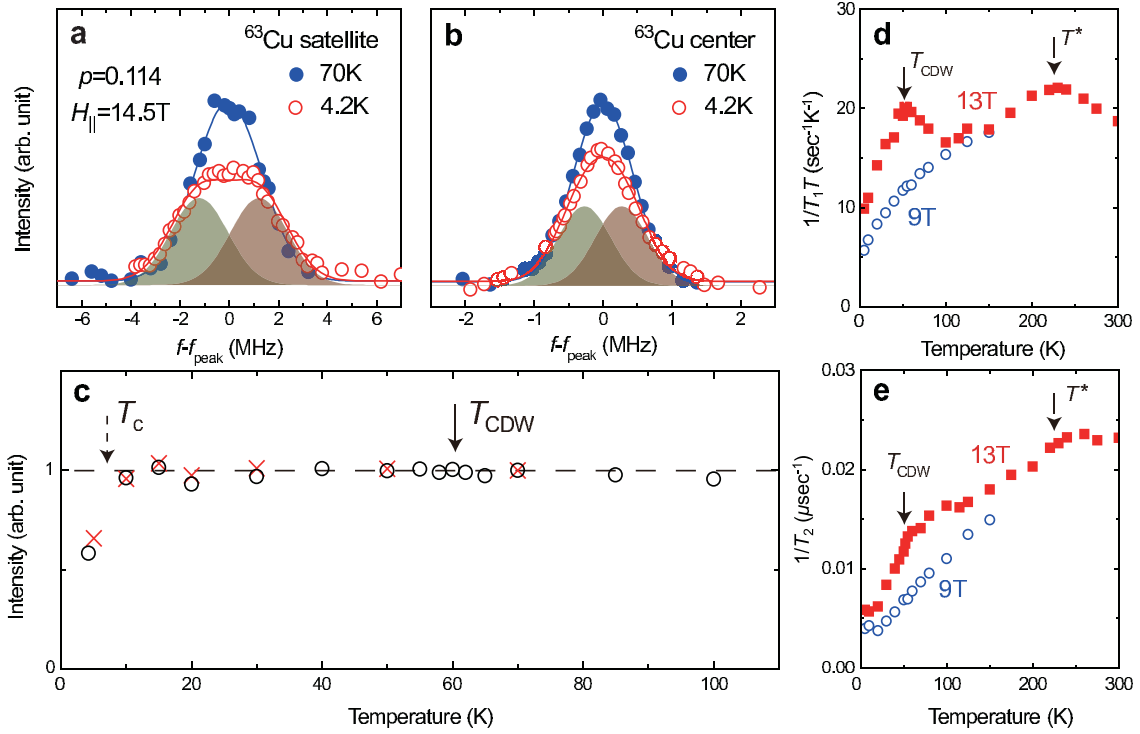}
        \end{center}
        \caption{Evidence for the CDW without a spin order. Temperature dependence 
        of the NMR satellite (a) and center (b) lines for $p$ = 0.114 at $H$ = 14.5 T. Solid 
        curves are the results of Gaussian fittings. The curves at $T$ = 4.2 K are the sum of 
        two Gaussian functions (shaded area). (c) Temperature dependence of NMR signal 
        intensity for the satellite line (open cicles) and the center line (crosses). The intensity is 
        corrected by taking into account  the 1/$T$ factor  and  the $T_2$ effect. The absence 
        of any intensity loss across $T_{\rm CDW}$ rules out the presence of any kinds of 
        spin order. The intensity loss below $T_{\rm c}$ is due to the Meissner  effect. The 
        temperature dependencies of $1/T_1T$ (d) and $1/T_2$ (e)  at different fields.} 
        \label{f2}
       \end{figure}

           \begin{figure}
                \begin{center}
                \includegraphics[width=11cm]{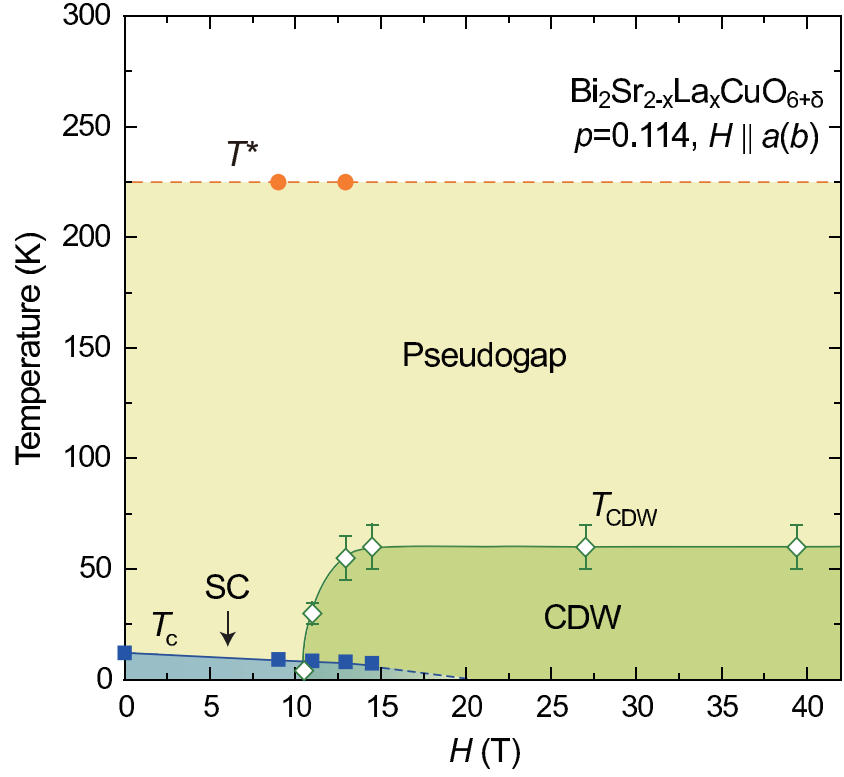}
                \end{center}
                \caption{$H$-$T$ phase diagram for underdoped Bi2201. The $T^*$ is 
                the 
                pseudogap temperature determined from the spin-lattice relaxation rate results. 
                Error 
                bars  represent the uncertainty in defining the onset temperature $T_{\rm 
                CDW}$.} 
           
                \label{f3}
                \end{figure}
              
  \begin{figure}
   \begin{center}
    \includegraphics[width=10cm]{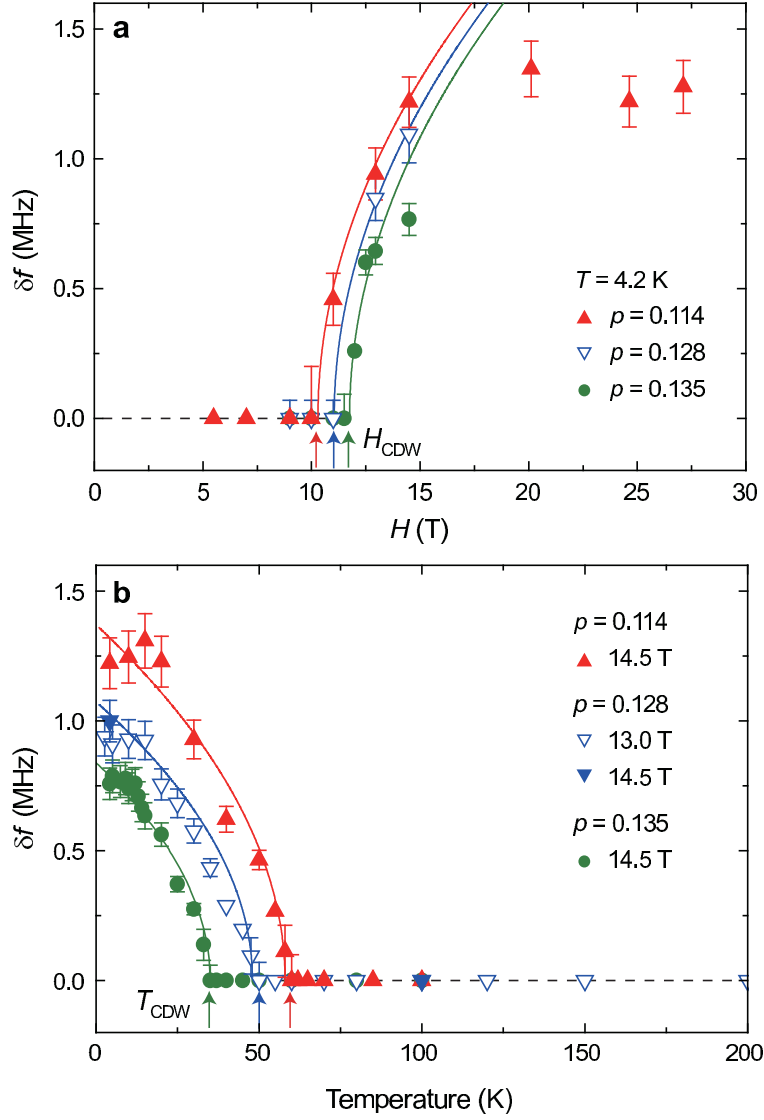}
   \end{center}
    \caption{Doping dependence of $H_{\rm CDW}$ and $T_{\rm CDW}$.  (a)  
    $H$ dependence of $\delta f$ at $T$ = 4.2 K. From fitting (solid curves)  the data to a 
    mean field model, $\delta f$ $\propto$ $(H-H_{\rm CDW})^{0.5}$, the threshold field 
    for $H$-induced CDW, $H_{\rm CDW }$= 10.4, 11.0, and 11.8 T for $p$ = 0.114,  
    0.128, and 0.135, respectively, was determined. (b) Temperature dependence of $\delta 
    f$ for the three samples, from which the CDW onset temperature $T_{\rm CDW}$ = 60, 
    50, and 35 K for $p$ = 0.114, 0.128, and 0.135, respectively, 
    	was determined. The curves are fits to a mean field model, $\delta f$ $\propto$ 
    	$(T_{\rm CDW}-T)^{0.5}$.  The dotted horizontal lines are guides to the eyes. Error 
    	bars represent the uncertainty in estimating $\delta{f}$.} 
     \label{f4}
     \end{figure}

      \begin{figure}
       \begin{center}
        \includegraphics[width=10cm]{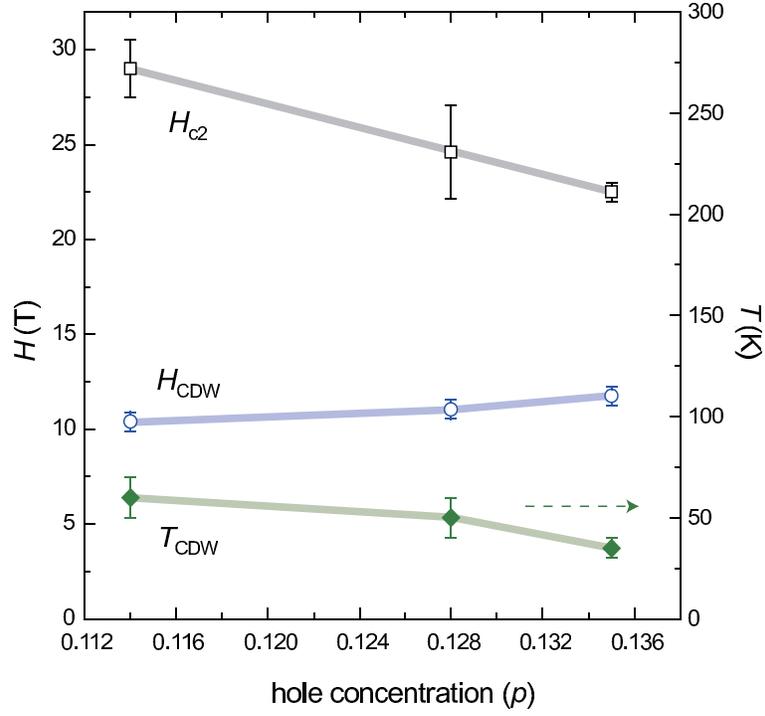}
       \end{center}
        \caption{The relationship between superconductivity and the CDW in 
        underdoped Bi2201. Doped hole concentration dependence of the upper critical field  
        $H_{\rm c2}$, $H_{\rm CDW}$ (left axis) and $T_{\rm CDW}$ (right axis). Error 
        bars represent the uncertainty in the fit using WHH formula to obtain $H_{\rm c2}$ 
        (Supplementary Fig. 5) and in defining the onset field $H_{\rm CDW}$ and 
        temperature  $T_{\rm CDW}$.} 
         \label{f5}
         \end{figure}

 \begin{figure}
	\begin{center}
		\includegraphics[width=13cm]{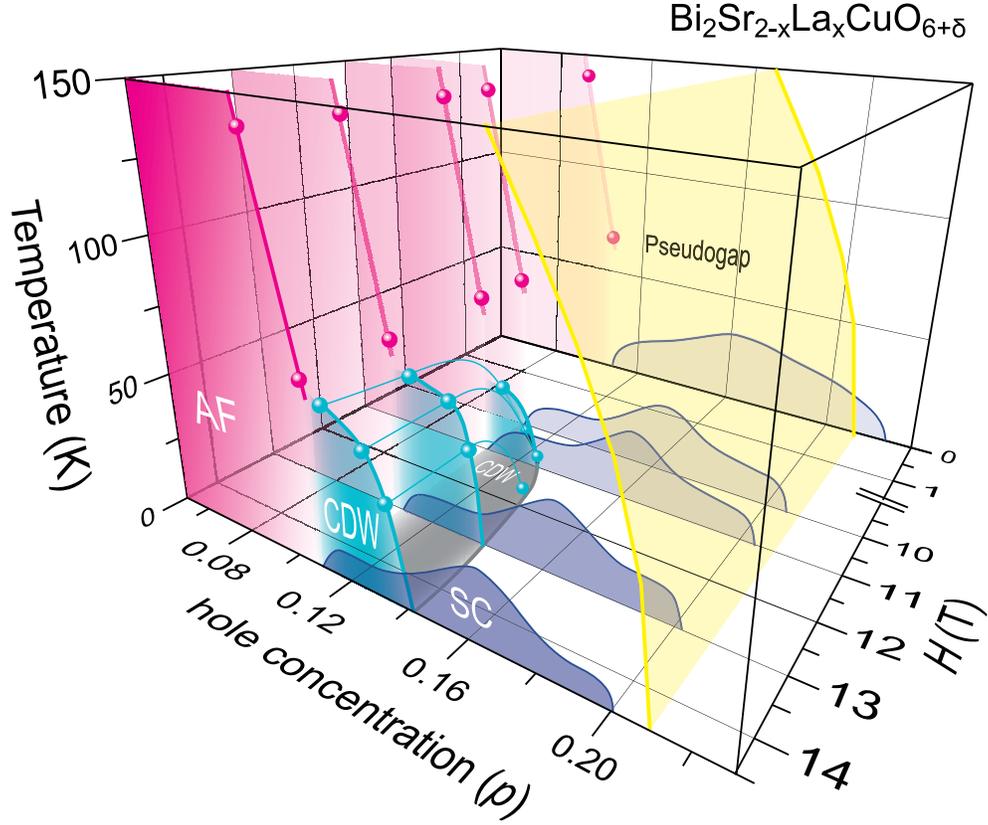}
	\end{center}
	\caption{Magnetic-field evolution of the phase diagram. Hole concentration ($p$) 
	dependence  of the pseudogap temperature $T^*$, $T_{\rm N}$ and $T_{\rm c}$, 
	and the $p$- and $H$-dependence of $T_{\rm CDW}$ for 
	Bi$_2$Sr$_{2-x}$La$_x$CuO$_{6+\delta}$. Magnetic field is applied along the Cu-O 
	bond direction ($H\parallel{a(b)}$). $T_{\rm CDW}(p,H)$ (green and gray curves) is 
	obtained from the results on Fig. 1f and Fig. 4.  $T^*$ (yellow curve) and   $T_{\rm 
	N}$ (red curve) are from the previous works \cite{Zheng,Kawasaki}. $T_{\rm c}(p,H)$ 
	(blue curve) is determined by the ac-susceptibility measurements using the NMR coil. }
	\label{f6}
\end{figure}

\begin{figure}
	\begin{center}
		\includegraphics[width=11cm]{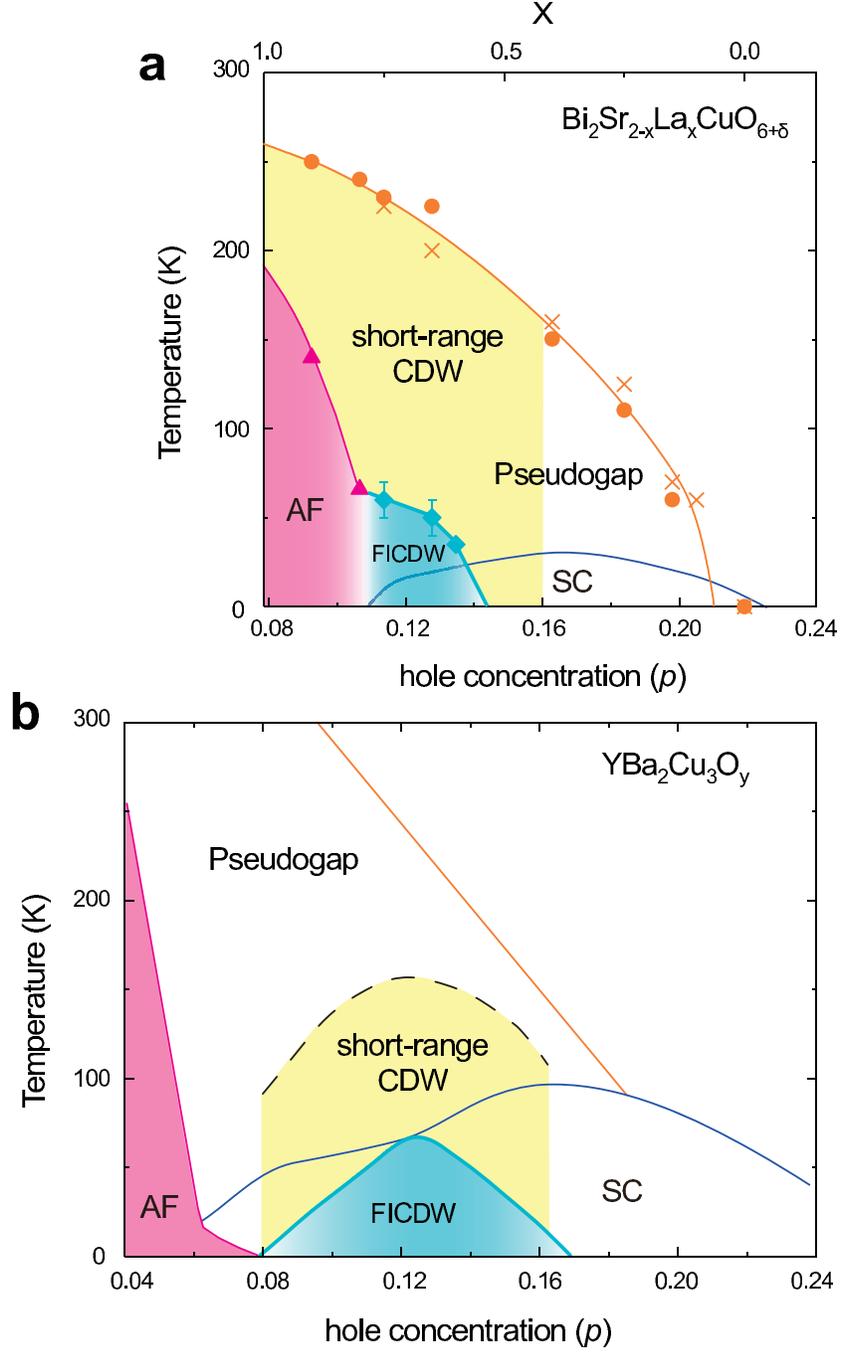}
	\end{center}
	\caption{Phase diagram of Bi2201 compared with YBCO.  Doping dependence of 
	the short-range CDW reported by the X-ray measurements  
	\cite{CominBi2201,PengYY,Uchidareview} and  the field-induced CDW (FICDW) for 
	Bi2201 (a) and YBCO (b). For both cases, $T_{\rm c}$ is the zero-field value.   Error 
	bars  represent the uncertainty in defining $T_{\rm CDW}$.  The schematic phase 
	diagram for YBCO is   from the literatures \cite{Uchidareview,WuNature}.   } 
	\label{f7}
\end{figure}

 \begin{figure}
 	\begin{center}
 		\includegraphics[width=10cm]{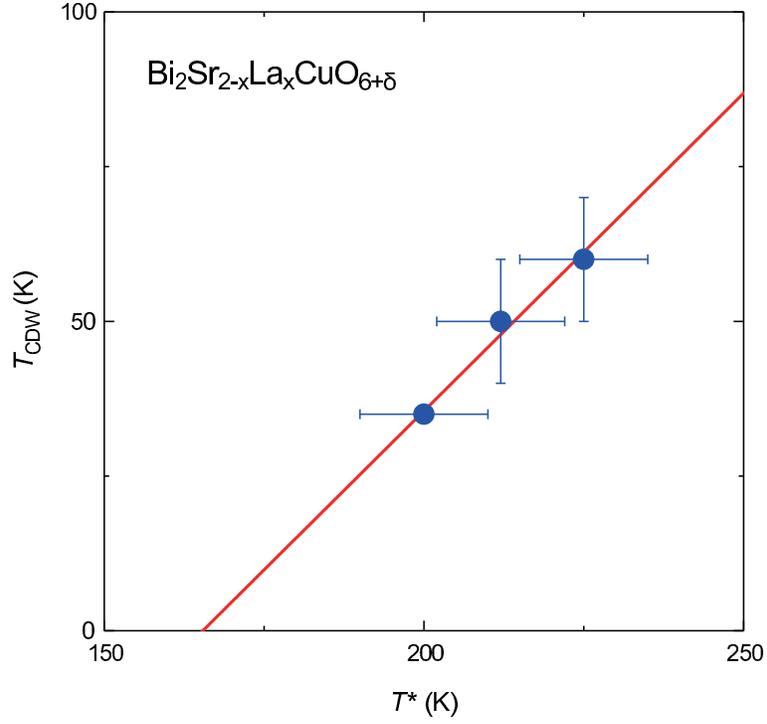}
 	\end{center}
 	\caption{Relationship between $T_{\rm CDW}$ and $T^*$.  A plot of  $T_{\rm 
 	CDW}$ vs $T^*$  at $H$ = 14.5 T. The straight line is a fitting to the data which yields a 
 	slope of 1.0 $\pm$ 0.7. Error bars represent the uncertainty in defining the two 
 	characteristic temperatures. } 
 	\label{f8}
 \end{figure}

\end{document}